\documentclass[reprint,amsmath,amssymb,aps,superscriptaddress]{revtex4-1}
\usepackage{dcolumn}
\usepackage{graphicx}
\usepackage{bm}
\usepackage{xcolor}

\begin{document}

\title{Novelty search in non-equilibrium systems}
\title{Novelty search in non-equilibrium systems}
\title{Learning to discover novel behaviors in non-equilibrium systems}
\title{Sampling for novel non-equilibrium behaviors in a continually updated order parameter space}
\title{Exploring non-equilibrium behaviors in a continually updated order parameter space}
\title{Novelty sampling of non-equilibrium behaviors with dynamically updated order parameters}
\title{Curiosity sampling of non-equilibrium behaviors using dynamically learned order parameters}
\title{Curiosity search for non-equilibrium behaviors in a dynamically learned order parameter space}
\title{Curiosity-driven search for novel non-equilibrium behaviors}




\author{Martin J. Falk*}
\affiliation{Department of Physics, The University of Chicago, Chicago, IL 60637}

\author{Finnegan D. Roach*}
\affiliation{Department of Physics, The University of Chicago, Chicago, IL 60637}

\author{William Gilpin}
\affiliation{Department of Physics, The University of Texas, Austin, TX 78712}

\author{Arvind Murugan}
\affiliation{Department of Physics, The University of Chicago, Chicago, IL 60637}

\begin{abstract}
Exploring the spectrum of novel behaviors a physical system can produce can be a labor-intensive task. Active learning is a collection of iterative sampling techniques developed in response to this challenge. However, these techniques often require a pre-defined metric, such as distance in a space of known order parameters, in order to guide the search for new behaviors. Order parameters are rarely known for non-equilibrium systems \textit{a priori}, especially when possible behaviors are also unknown, creating a chicken-and-egg problem. Here, we combine active and unsupervised learning for automated exploration of novel behaviors in non-equilibrium systems with unknown order parameters. 
We iteratively use active learning based on current order parameters to expand the library of known behaviors and then relearn order parameters based on this expanded library.
We demonstrate the utility of this approach in Kuramoto models of coupled oscillators of increasing complexity. In addition to reproducing known phases, we also reveal previously unknown behavior and the related order parameter.
\color{black}Finally, we demonstrate how curiosity-driven search can naturally be aligned with human intuition.\color{black}
\end{abstract}

\maketitle

When handed a new experimental platform, our first instinct is to go exploring - to tune individual experimental parameter knobs and record the resulting behaviors of the system. This scattershot investigation provides a window into the range of behaviors that can be produced. In this way, we build intuition for the right variables to describe the system which in turn can serve as a prelude for more systematic quantitative investigation.

However, as the experimental parameters we have access to grow increasingly high-dimensional (e.g. space- and time-dependent activity\cite{ross2019controlling,volpe2011microswimmers,buttinoni2012active,zhang2021spatiotemporal} or interactions in many-body active systems\cite{bauerle2018self,wang2021emergent}), and the resulting behaviors grow increasingly complex, it becomes a labor-intensive task to explore the full spectrum of behaviors.
\color{black}Brute-force approaches such as grid search quickly become prohibitive even in dimensions where optimization would be feasible. \color{black}
Much of the parameter space may be uninteresting, and in the absence of previously built intuition or an analytical theory, it is difficult to know which parts of parameter space might show useful or novel behaviors.
Hence we are faced with a twinned challenge; how to efficiently search the space of experimental parameters to reveal novel behaviors, while also learning to characterize the behaviors in terms of order parameters.


Individually, these problems have been recognized and addressed in creative ways. 
On the parameter side of the challenge, active learning\cite{dai2020efficient,whitelam2021neuroevolutionary,ferguson2022data,shmilovich2020discovery,mohr2022data,grizou2020curious,oudeyer2007intrinsic,reinke2019intrinsically} provides iterative methods for efficiently sampling parameter spaces.
In these approaches, behaviors collected at previously sampled parameters inform parameter sampling in the future, so as to increase the likelihood of discovering novel behaviors. However, these techniques require a metric in the space of behaviors, which often takes the form of a distance in a space of known order parameters. For most non-equilibrium many-body systems, such order parameters are not known. 

However, to find order parameters, one needs to know the range of possible behaviors, thus creating a chicken and egg problem. 
On the behavior side of the challenge, 
data-driven dimensionality reduction techniques\cite{coli2022inverse,thiem2020emergent,carrasquilla2017machine,mcgibbon2017identification,van2020classifying,vaddi2022autonomous,gilpin2021chaos,ricci2022phase2vec} can reveal a small number of order parameters from a library of known behaviors.  
But these methods require a sufficiently comprehensive library of behaviors to infer meaningful order parameters.



Here, we will demonstrate how a curiosity-driven search algorithm can efficiently explore non-equilibrium many-body systems, even in the absence of previously known order parameters. 
We adapt methods that combine the strengths of both active learning and dimensionality reduction\cite{oudeyer2007intrinsic,baranes2013active,reinke2019intrinsically}. 
We learn order parameters through unsupervised dimensionality reduction on a library of currently known behaviors; we then use active learning in the space of current order parameters to reveal new behaviors and iterate. 
\color{black}Crucially, we always search in the learned low dimensional latent space \textit{trained on dynamical behaviors} rather than the high dimensional parameter space; in this way, active learning efficiently samples richer parts of parameter space.\color{black}




\begin{figure}
\begin{centering}
\includegraphics[width=\linewidth]{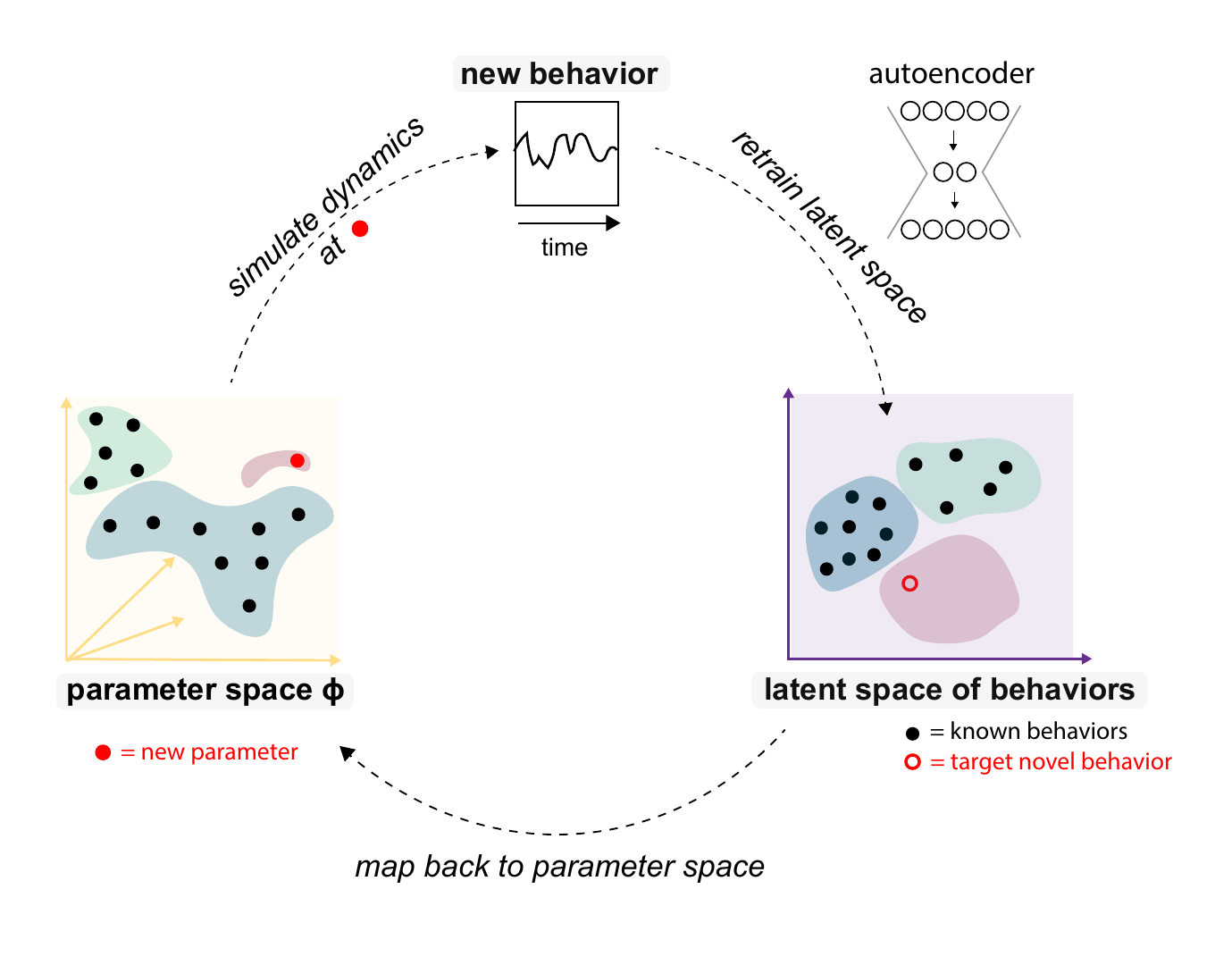}
\par\end{centering}
\caption{Overview of the curiosity-driven search for novel behaviors.
We consider a system with a high dimensional parameter space (yellow) whose potential behaviors and order parameters that might describe them are initially unknown. Search is initialized by collecting behaviors corresponding to a uniform sampling of parameter space. These dynamical behaviors are used to train an autoencoder to obtain a low dimensional latent space of behaviors (purple) parameterized by putative order parameters. We then seek a new behavior (`curiosity') by randomly sampling the learned latent space (open red circle) rather than sampling parameter space. We map the target new latent space point back to parameter space (solid red circle), evaluate the resulting behavior and thus expand our library of known behaviors. 
Autoencoder is retrained every K training rounds on a random subset of previously sampled behaviors, thus improving the learned latent space and order parameters. (Green, purple and blue regions of parameter and latent spaces indicate qualitatively distinct behaviors.)
}
\label{fig:algorithm_schematic}
\end{figure}

We apply our general framework to a paradigmatic class of dynamical systems - the Kuramoto model of oscillators and their variants\cite{kuramoto1975self,acebron2005kuramoto}.
We first use curiosity search to recapitulate known results on simple Kuramoto model variants with one or two parameters, which are nevertheless capable of producing rich non-equilibrium behaviors.
We then explore a 3-population Kuramoto model with 10 adjustable parameters and reveal previously un-characterized behavior and corresponding new order parameters.
\color{black}Finally, we demonstrate how curiosity search can be formulated to naturally align with human intuition in order to target multi-population behaviors in a 10-population Kuramoto model with 100 parameters.\color{black}

Our work establishes a general framework that can be used with other models of complex systems or can directly interface with an experimental system where no model is available. 



\section*{Method}

\begin{figure*}
\begin{centering}
\includegraphics[width=\linewidth]{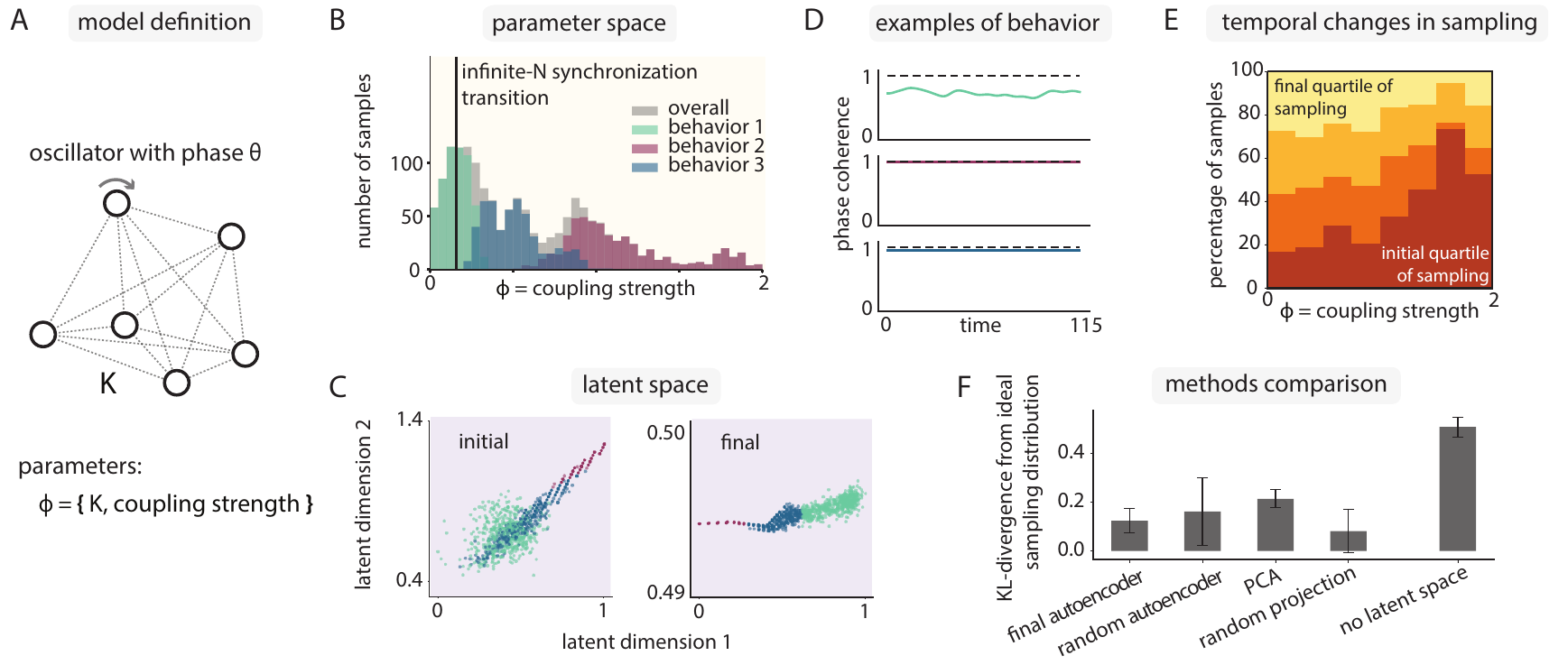}
\par\end{centering}
\centering{}\caption{Curiosity search reveals all known phases and order parameters for the uniformly connected Kuramoto model more efficiently than random search.
(A) In the canonical Kuramoto model, $N$ oscillators are coupled to favor alignment (coupling strength $K$)
Here number of oscillators $N = 33$.
(B) Unlike a random search of $K$ space, our algorithm samples non-uniformly, focusing on the less common desynchronized state at small $K$. Vertical line indicates phase boundary $K_c$ computed for $N \to \infty$. Behaviors and the associated colors are computed from latent space.
(C) Autoencoder latent space at the start and end of curiosity search; the final latent space identifies a quasi-1-dimensional structure for oscillator behavior, indicating one useful order parameter. Clustering in latent space categorizes collective behaviors, corresponding to distinct regions of parameter space. 
(D) Phase coherence examples from dynamical states identified through latent space clustering.
(E) Curiosity search increasingly focuses on sampling lower $K$ as training proceeds. 
(F) Curiosity search works with other dimensionality reduction methods, consistently generating better parameter sampling than random sampling (no latent space). Error bars indicate variance over 10 replicates.}
\label{fig:fullyconnected_figure}
\end{figure*}

The curiosity sampling algorithm has three key components shown in Fig. \ref{fig:algorithm_schematic}: a potentially high-dimensional parameter space (yellow); a potentially high-dimensional space of raw system behaviors; and a lower-dimensional latent space of behaviors (purple).
Our algorithm is as follows:  We initialize by randomly sampling parameter space, and compile the corresponding library of behaviors by integrating the equations of motion for these parameter choices. We then train a dimensionality reduction method on the library of behaviors assembled so far through parameter space exploration, revealing an updated latent space of behaviors and order parameters.  Then, crucially, we search for new behaviors in this emergent latent space of behaviors created by dimensionality reduction. The new target behaviors are then mapped back to a new point to sample in parameter space. We evaluate the behavior for these parameter choices by integrating the equations of motion, thereby expanding our library of known behaviors. Finally, after a certain number of new parameters are sampled, we retrain the dimensionality reduction and the cycle repeats. 

We emphasize that the goal of this algorithm is to construct a space which captures different possible behaviors of the underlying physical system. Intuitively, the latent behavior space is a more efficient space for sampling than parameter space or the full space of behaviors with an arbitrary metric, as the latent space represents the most relevant aspects of behavior. 
Additionally, sampling in the latent space of behaviors can up-weight behaviors that are rare in parameter space but constitute a significant region of a phase diagram.

\color{black}The key feature of the curiosity-driven search is therefore the way latent spaces are constructed from the timeseries output of the dynamical system, and \textit{not} the control parameter space.\color{black}

Our algorithm, outlined in general above, has several choices in the details of how different steps are implemented. The mapping from dynamical behavior to latent space involves both pre-processing and dimensionality reduction.
In our results, we will preprocess by mean-centering and binning sampled time-points to account for permutation invariance and global mean rotations.
For dimensionality reduction, we use a convolutional variational autoencoder\cite{kingma2013auto} (VAE) with relatively simple encoder and decoder architectures.
In part to guard against cherry-picking model architectures, we also compare the results of the VAE-based dimensionality reduction to other non-neural net methods. See Appendix \ref{dimreduction_details} for further details.

Additionally, the sampling and backmapping of latent space points to parameter space can occur through several different methods.
In what follows, we choose a particularly simple implementation of the back-mapping; when sampling a new latent space goal, we look to the nearest previously sampled latent space point, and identify its associated parameter space point.
We can then make a random step from this nearest-neighbor parameter space point.
In this way, we make a guess at what points in parameter space are likely to produced a dynamical behavior with our targeted latent space goal.
Our choice for latent space sampling is similarly simple; we uniformly sample the bounding hypercube of the current set of collected latent space points.
For further details, see Appendix \ref{active_learning_details}, and Limitations and Extensions for a discussion of other latent space and backmapping methods.


\section*{Results}

To evaluate the performance of a curiosity-driven search in a well-characterized setting, we turn to the original formulation of the Kuramoto model\cite{kuramoto1975self},
\begin{equation}\label{kuramoto_model_original}
\dot{\theta_i} = \omega_i  +  \frac{K}{N}\sum_{j=1}^N \sin(\theta_i - \theta_j),
\end{equation}
\noindent where the $\omega_i$ are drawn independently from a distribution $\mathcal{N}(0,.1)$, and the coupling strength $K > 0$ is the one tunable parameter (Fig. \ref{fig:fullyconnected_figure}A). We set $N = 33$ for our simulations.

In the limit of infinite $N$, this model is characterized by a critical coupling strength\cite{acebron2005kuramoto} $K_c=.16$ for our parameters.
For values of $K < K_c$, the oscillators move independently of each other, creating a desynchronized behavior.
For values of $K > K_c$, the oscillators synchronize and have exactly the same phase $\theta$.

Let's pretend that we are approaching this system without prior knowledge about the behaviors that can arise, and where these transitions occur.
In other words, the only information we have about the system is that there is one parameter that we can manipulate, which is $K$.
We will make the assumption that interesting behaviors occur in the system when the coupling strength is $O(1)$ or less.
One way to approach exploration of this system would be to randomly sample values of $K$, and observe the behavior at these sampled values.
With this approach, only a small fraction of the observed behaviors would be desynchronized, since $K_c$ is $O(.1)$.


\begin{figure*}
\begin{centering}
\includegraphics[width=\linewidth]{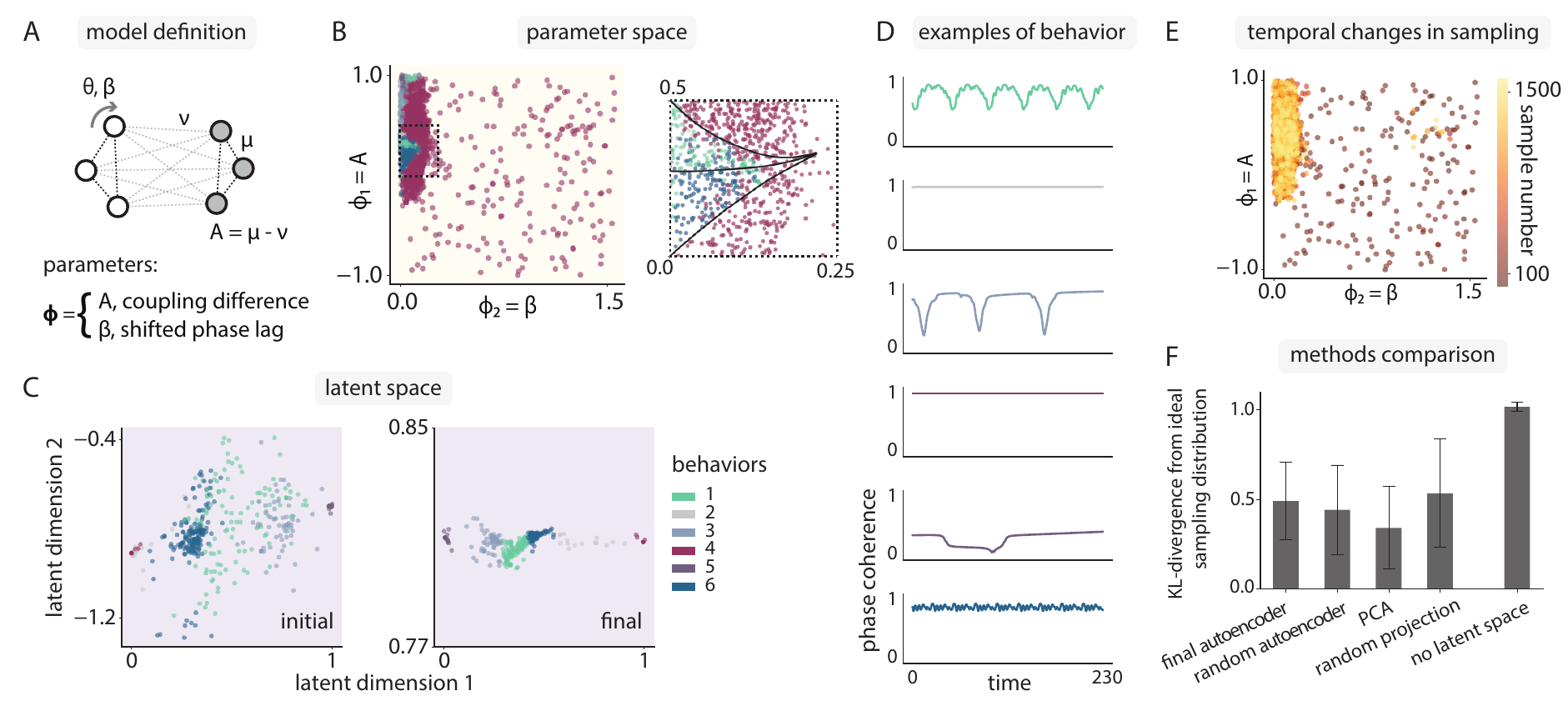}
\par\end{centering}
\centering{}\caption{Curiosity search efficiently reveals the full phase diagram for a 2-population Kuramoto model. (A) Kuramoto model with two populations of oscillators considered by Abrams \emph{et al}\cite{abrams2008solvable}, with intra-population coupling $\mu$, inter-population coupling $\nu$ and shifted phase offset $\beta$, with number of oscillators $N = 32$.
(B) The curiosity search samples non-uniformly across parameter space, focusing on the small region where rare chimera behaviors occur. (inset) Clustering in latent space reveals that this region has the structure of the chimera stability diagram identified by Abrams \emph{et al}\cite{abrams2008solvable}.
(C) Autoencoder latent space at the start and end of curiosity search; desynchronized states occupy more space relative to synchronized states than in the parameter space.
(D) Phase coherence examples from each of the states identified through latent space clustering.
(E) As training proceeds, curiosity search increasingly focuses on parameter space where desynchronized states are found.
(F) Curiosity search works with other dimensionality reduction methods, consistently generating better parameter sampling than random sampling (no latent space). Error bars indicate variance over 10 replicates.}
\label{fig:chimera_figure}
\end{figure*}

Running our curiosity search in the one-dimensional parameter space of coupling strength demonstrates the features of successful system exploration.
In the final ensemble of collected parameters, samples are drawn with frequencies weighted towards couplings of $O(.1)$, where we expect the infinite-N synchronization to occur (Fig. \ref{fig:fullyconnected_figure}B).
We can interpret this weighted sampling as the curiosity search algorithm having learned to distinguish the synchronized and desychronized phases.
The fully-trained latent space also provides evidence for ``learning'' of the Kuramoto model behaviors, as the final latent space is a thin 1D manifold with the same ordering as the parameter space.
Clustering by agglomerative clustering, as a post-data collection step, can readily reveal this ordering by showing how contiguous regions of latent space are mapped to parameter space (Fig. \ref{fig:fullyconnected_figure}C).
By plotting as a summary statistic the traditional Kuramoto phase coherence $|\frac{1}{N}\sum_{j=1}^N e^{i\theta_j}|$, we see that individual examples of the dynamical behaviors confirm this picture as well (Fig. \ref{fig:fullyconnected_figure}D), revealing desynchronized (behavior 1), synchronized (behavior 2), and intermediate (behavior 3) behaviors.
Finally, we see that sampling bias towards the desynchronized region increases as sampling progresses, indicating that the curiosity search is changing its latent space over time to better reflect the relevant behaviors (Fig. \ref{fig:fullyconnected_figure}E).

To test whether other algorithms could have performed the same task, we considered multiple variants of the dimensionality reduction technique: PCA, a random autoencoder which was never trained, and a random linear projection (see Appendix \ref{dimreduction_details} for further detail).
As we have access to a prior understanding of the dynamical behaviors present in the model, we can post-collection compare the distribution of sampling to an ideal distribution which samples the known behaviors equally (Fig. \ref{fig:fullyconnected_figure}F).
All dimensionality reduction consistently outperformed random sampling of parameter space.
We note the surprising result that random projection outperformed even iteratively trained methods.
The success of random methods parallels observations made in the context of timeseries featurization with random convolutions\cite{dempster2020rocket}.
It also indicates that there was enough structure already present in the raw dynamical systems output such that a random low-dimensional projection was able to separate the various accessible behaviors.

\begin{figure*}
\begin{centering}
\includegraphics[width=\linewidth]{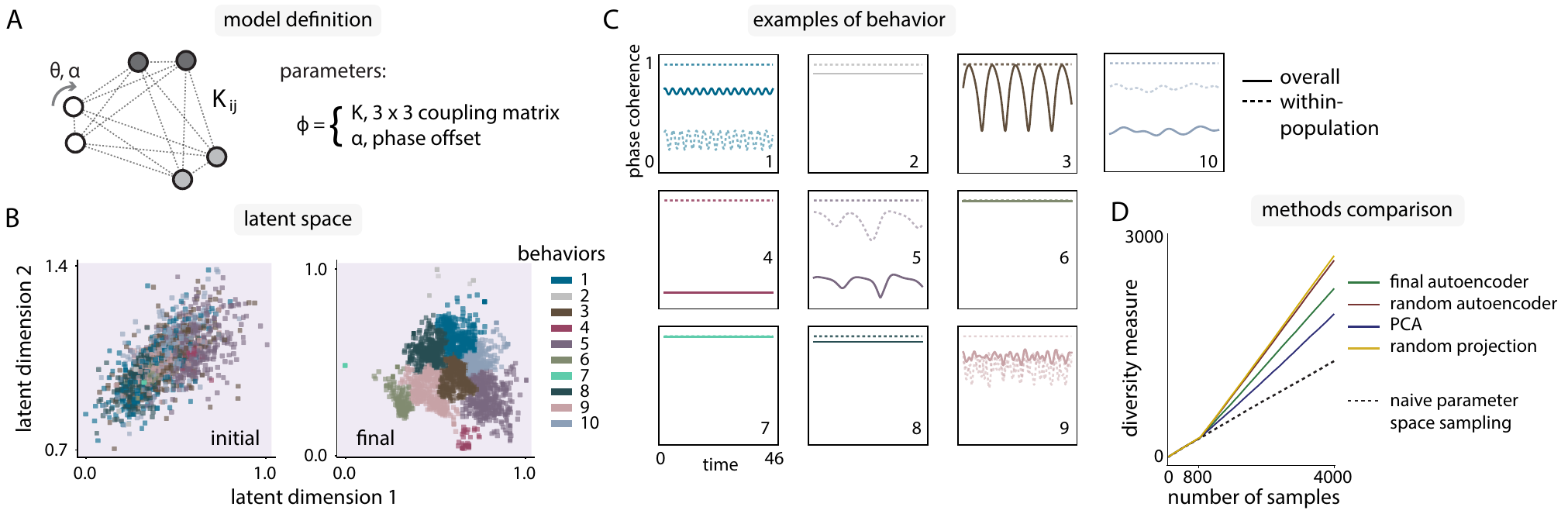}
\par\end{centering}
\centering{}\caption{Curiosity search in a 3-population Kuramoto model reveals a previously unknown phase, order parameters and transferable insights.
(A) A 3-population Kuramoto model with 9 population-population couplings, and 1 global phase offset $\alpha$. Couplings $K_{ij}$ are positive, and number of oscillators $N = 30$. (B) Autoencoder latent space at the start and end of curiosity search.
(C) Phase coherence examples from each of the states identified through latent space clustering. Solid lines represent overall phase coherence, dashed lines are phase coherence of indiviual populations.
(D) All dimensionality reduction variants of the curiosity search algorithm generate more diversity than random sampling (black dashed line). Each line is computed from 10 replicates. See SI for details on diversity measure calculation.}
\label{fig:3population_figure}
\end{figure*}

While the uniformly-connected Kuramoto model is an ideal testing ground, the range of dynamical behaviors it can produce is fairly simple. 
We extend our approach to a Kuramoto model variant whose phase diagram has been equally well-characterized, but is capable of producing a wider range of behaviors, including chimera states.

Specifically, we investigate a 2-population Kuramoto model with a coupling $K_{11} = K_{22} = \mu$ between all oscillators within the same population, and a coupling $K_{12} = K_{21} = \nu$ between all oscillators in different populations.
Subscripts indicate the oscillator population index.
We introduce a phase offset $\alpha$ to the coupling between any two oscillators and write the model as:

\begin{equation}\label{kuramoto_model_chimera}
\dot{\theta_i^\sigma} = \omega  +  \sum_{\sigma'=1}^2 \frac{K_{\sigma\sigma'}}{N_{\sigma'}} \sum_{j=1}^{N_{\sigma'}} \sin(\theta_j^{\sigma'} - \theta_i^\sigma - \alpha),
\end{equation}

This model was introduced by Abrams \emph{et al}\cite{abrams2008solvable}, where the parameter space was given by the variables $\beta = \frac{\pi}{2} - \alpha$ and $A = \mu-\nu$ (Fig. \ref{fig:chimera_figure}A), with $\mu+\nu=1$. Here, we specifically investigate the case $\omega = 0$ and total $N = 32$, with equal population sizes. We will term this model the ``chimera’’ model, as it was shown to produce chimera states, where two identical populations of oscillators exist with one synchronized and the other desynchronized\cite{cho2017stable,nicolaou2019multifaceted,zhang2020critical,abrams2004chimera}. 

Employing curiosity search in this 2-dimensional parameter space results in a distribution of samples that is concentrated on a narrow strip of the total parameter space, roughly in the area with $A > 0$ and $\theta < .25$ (Fig. \ref{fig:chimera_figure}B). 
This is precisely the region of parameter space which is known to support the emergence of chimeric behavior. 
In fact, the latent space trained through our curiosity sampling procedure is able to distinguish between the two types of chimeras originally identified by Abrams \emph{et al}\cite{abrams2008solvable} (Fig. \ref{fig:chimera_figure}B(inset), C), despite the fact that our analysis is done on a smaller number of oscillators, and those results were derived in an infinite-N limit.


Visualizations of the dynamical behaviors provide additional evidence that automated curiosity sampling is capturing a wide variety of behaviors in the chimera model (Fig. \ref{fig:chimera_figure}D), and the temporal changes in sampling indicate that the parameter regions which contain the richest dynamical behaviors are preferentially sampled as the latent space is trained (Fig. \ref{fig:chimera_figure}E).
Behaviors 1 and 6 correspond to the breathing and stable chimeras respectively.
As we have access to a prior understanding of some of the dynamical behaviors present in the model, we can post-collection compare the distribution of sampling to an estimated ideal distribution which samples the known behaviors equally (Fig. \ref{fig:chimera_figure}F).
All dimensionality reduction consistently outperformed random sampling of parameter space.

Having investigated the utility of automated curiosity sampling in a non-trivial but still thoroughly explored model, we now turn to previously unexplored models.
We initially define a 10-dimensional variant of the chimera model, with three populations with phase offset (Fig. \ref{fig:3population_figure}A):

\begin{equation}\label{kuramoto_model_3population}
\dot{\theta_i^\sigma} = \omega  +  \sum_{\sigma'=1}^3 \frac{K_{\sigma\sigma'}}{N_{\sigma'}} \sum_{j=1}^{N_{\sigma'}} \sin(\theta_j^{\sigma'} - \theta_i^\sigma - \alpha),
\end{equation}
\noindent with $\omega = 0$ and total $N = 30$ divided equally among individual populations.
The coupling matrix between the populations is not restricted to be symmetric, though we require all matrix elements to be positive.

Due to the 10-dimensional nature of the parameter space, we forego the direct visualization of parameters and instead focus on visualizing our (4-dimensional) latent space.
We select the two dimension in latent space which contribute the most to the largest principal component of the trained latent space, and project our data on these axes (Fig. \ref{fig:3population_figure}B).
Clustering of behaviors in latent space shows that the latent space structure significantly changes between initial and final rounds of sampling.

To understand the behavior regimes in this latent space, we can visualize representatives of each group for qualitative analysis; both the overall phase coherence as well as the phase coherence of each individual population (Fig. \ref{fig:3population_figure}C).
We find a variety of behaviors, most of which can be interpreted in the light of previous behaviors uncovered in Kuramoto models -- fully synchronized\cite{kuramoto1975self} (behaviors 6, 7), chimera\cite{abrams2008solvable} (behaviors 1, 9), chiral\cite{fruchart2021non} (behaviors 2, 8), antialigned\cite{fruchart2021non} (behavior 4), and combination chiral + chimera phases (behaviors 5, 10).

Finally, to conclude our automated analysis of the 3-population Kuramoto model, we quantitatively confirm the relative diversity of samples compared to a random sampling baseline (Fig. \ref{fig:3population_figure}D).
The collapse of all curves below 800 samples corresponds to the initial random sampling.
In contrast to the uniformly-connected and chimera models, we lacked any prior knowledge of the phase behavior in parameter space.
We therefore adopted a model-agnostic measure of diversity corresponding to the total volume of trained autoencoder latent space occupied by another sampling distribution.
See Appendix \ref{sec:diversity_measure} for further detail.

\begin{figure}
\begin{centering}
\includegraphics[width=\linewidth]{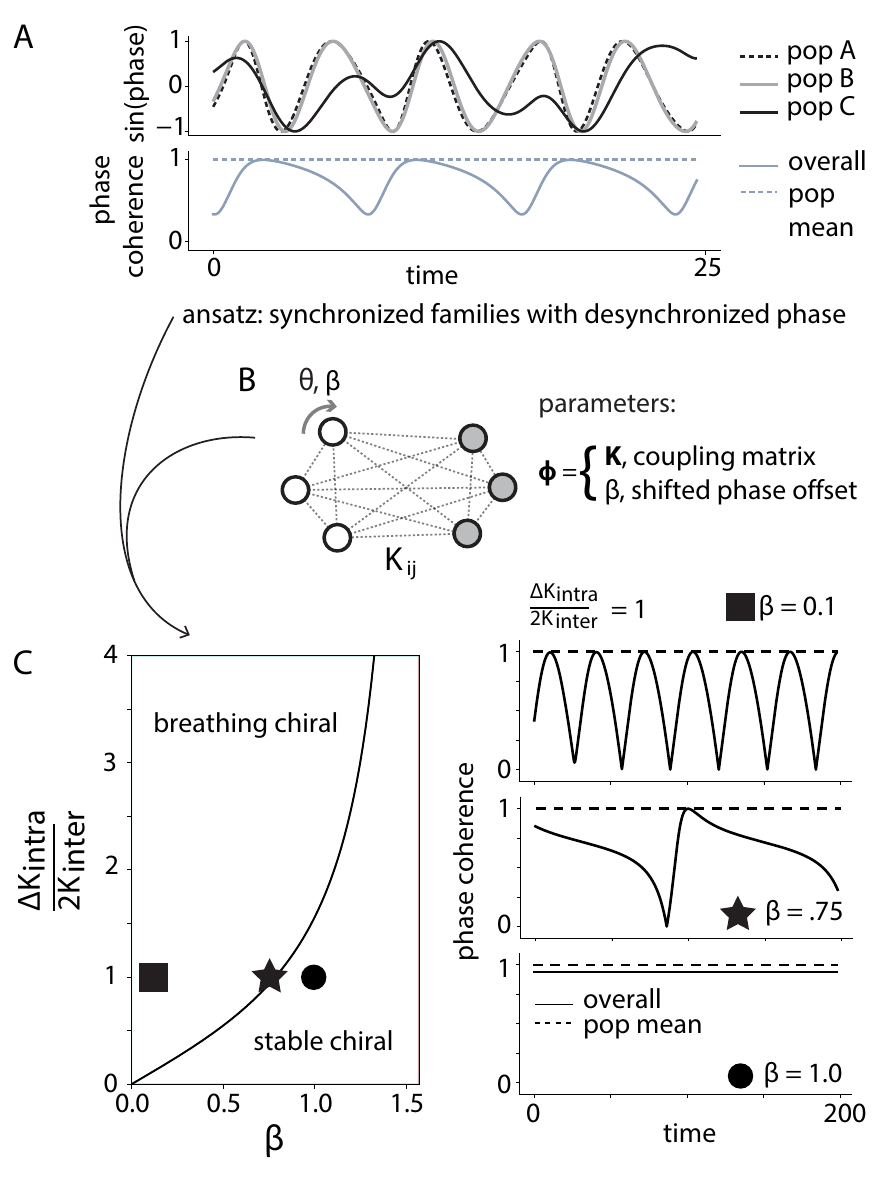}
\par\end{centering}
\centering{}\caption{Automated curiosity search in a 3-population Kuramoto model yields transferable insights for other models.
(A) One of the behaviors found in the curiosity search shows complete synchronization within populations, but with one population desynchronized from the other two: a ``chiral breather'' state. The dashed population mean indicates the average of the 3 phase coherence curves for the individual populations, with number of oscillators $N = 30$. 
(B) This behavior can be used as an ansatz for solving a simpler two-population Kuramoto model (infinite-N limit), with coupling matrix $K_{ij}$ and shifted phase offset $\beta$.
(C) Phase diagram of breathing and stable chiral states as a function of population couplings and phase offset (left). Dynamics of overall and population mean phase coherence at specific points in phase diagram, with number of oscillators $N = 32$ (right).}
\label{fig:newbehavior_fig}
\end{figure}

In our exploration of the 3-population Kuramoto model, we identified a particular set of parameters that led to an unexpected behavior (Fig. \ref{fig:3population_figure}C, behavior 3), where the phase coherence of each individual oscillator family was saturated, but the overall phase coherence displayed periodic variability.
We were particularly interested in understanding this behavior, as it did not neatly fit into any categories that we had previously encountered, resembling a chiral phase identified in Fruchart \emph{et al}\cite{fruchart2021non}, but with periodic breathing.

We took a closer look at these ``chiral breather'' dynamics, and found that the behavior came as a result of 2 populations completely synchronizing with each other, while a third population internally synchronised but moved at a different period relative to the other populations (Fig. \ref{fig:newbehavior_fig}A). 

In order to understand the chiral breather, we used the ansatz of internally synchronized families with a externally desynchronized phase to identify its emergence in a simpler system.
We chose to investigate a 2-population version of the 3-population model (Fig. \ref{fig:newbehavior_fig}B), which is identical to Eq. \ref{kuramoto_model_chimera}, without the inter- and intra-population coupling symmetry assumptions.

\begin{figure*}
\begin{centering}
\includegraphics[width=\linewidth]{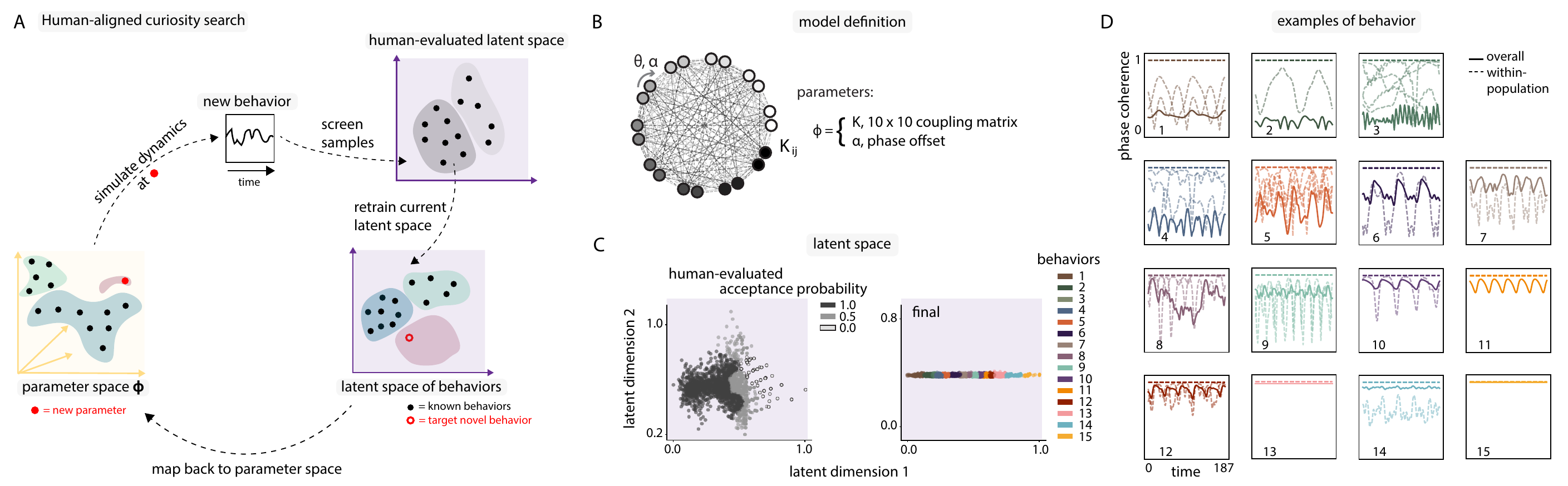}
\par\end{centering}
\centering{}\caption{\color{black}Human-aligned curiosity search discovers chaotic multi-population behaviors in 100-dimensional Kuramoto model.
(A) Curiosity search can naturally be aligned with human intuition with a single additional step, involving freezing the latent space of an initial curiosity searcher. Behavior samples drawn for subsequent curiosity search can be screened through the initial, frozen latent space, where samples are accepted or rejected based on probabilities assigned to different parts of latent space by a human observer. Following screening through this human-evaluated latent space, sampling proceeds as Fig. \ref{fig:algorithm_schematic}, with a separate autoencoder for the new, human-aligned latent space.
(B) A 10-population Kuramoto model with 100 population-population couplings, and one global phase offset $\alpha$. Couplings $K_{ij}$ are positive and sum to 1, with number of oscillators $N = 100$.
(C) (left) Autoencoder latent space following initial curiosity search without human alignment. Latent space clusters identified in initial search are human-evaluated for interest and assigned acceptance probabilities. (right) Final latent space at the end of human-aligned curiosity search.
(D) Phase coherence examples from each of the states identified through latent space clustering. Solid lines represent overall phase coherence, dashed lines are phase coherence of indiviual populations.}
\label{fig:10population_figure}
\end{figure*}

Following the procedure outlined for the chimera model by Abrams \emph{et al}\cite{abrams2008solvable}, we derive a set of coupled differential equations for the phase difference and coherence of the two oscillator populations in the limit of infinite population size.
Using our ansatz inspired from our data-driven exploration in Fig. \ref{fig:3population_figure}, we compute the steady-state behavior of the oscillators as a function of the model parameters (Fig. \ref{fig:newbehavior_fig}C(left)).

To derive these equations, we can take as a starting point Eq. (9) in Abrams \emph{et al}\cite{abrams2008solvable}, which is:
\begin{equation}\label{chimera_starting}
\begin{aligned}
    0 = {} & \dot{a_1} +  \frac{1}{2} a_1^2(K_{11}a_1^* + K_{12}a_2^*)e^{-i\alpha} \\
     & - \frac{1}{2} (K_{11}a_1^* + K_{12}a_2^*)e^{i\alpha},
\end{aligned}
\end{equation}
\noindent with the equation for $\dot{a_2}$ being identical under the interchange of subscripts 1 and 2. 
The $a_i$ are the amplitudes of the remarkable Ott-Antonsen ansatz\cite{ott2008low} for the oscillator phase density in the $N \to \infty$ limit.
Unlike in Abrams \emph{et al}\cite{abrams2008solvable}, we do not yet make any assumptions on the $K$s.

In Abrams \emph{et al}\cite{abrams2008solvable}, the amplitudes $a_i$ are rewritten in polar form, with $a_i = \rho_i e^{-i \phi_i}$.
However, because we are instead interested in the behavior exhibited in Fig. \ref{fig:newbehavior_fig}A, we make a different ansatz, and assume that $\rho_i = 1$ for both $i$. In this case, Eq. \ref{chimera_starting} reduces to:

\begin{equation}\label{chimera_starting_reduced}
    0 = {} \dot{\phi_1} +  K_{11} \sin \alpha  + K_{12}\sin (\alpha + \phi_1-\phi_2),
\end{equation}
\noindent with the associated equation for index 2 simply involving the exchange of subscripts for 1 and 2. We can define $\psi = \phi_1 - \phi_2$, in which case we have one equation
\begin{equation}\label{chimera_final}
\begin{aligned}
    \dot{\psi} = {} & - [(K_{11} - K_{22})\sin\alpha + K_{12}\sin(\alpha + \psi) \\ 
    & - K_{21}\sin(\alpha-\psi) ].
\end{aligned}
\end{equation}
\noindent Integrating yields
\begin{equation}\label{chimera_solution}
\begin{aligned}
    \psi(t) = {} & 2 \tan^{-1}\left[\frac{D\tan(-\frac{Dt}{2\sqrt{2}}+c_0)-A}{B}\right]\\
    A = {} & (K_{12}-K_{21})\cos\alpha,\\
    B = {} & \sqrt{2}\sin\alpha((K_{11} - K_{22})-(K_{12} - K_{21}))\\
    D = {} & ((K_{11}-K_{22})^2 - 2(K_{12}^2 + K_{21}^2) \\ & - ((K_{11}-K_{22})^2 + 4 K_{12}K_{21})\cos2\alpha )^\frac{1}{2},
\end{aligned}
\end{equation}
\noindent where $c_0$ is a constant of integration. 

We note that there are two behaviors embedded in this solution, depending on $\operatorname{Im}(D)$.
When $D$ is real, $\psi$ continues to change over time as $t \to \infty$, indicating a chiral breather.
If $D$ is imaginary, then because of the conversion between $\tan$ and $\tanh$, $\psi$ goes to a constant in the long-time limit.
In the case where $K_{12} = K_{21} = K_{inter}$ and we define $\Delta K_{intra} = (K_{11}-K_{22})^2$, the boundary between these two behaviors simplifies to:
\begin{equation}\label{bound}
    \Delta K_{intra}^2 = 4 K_{inter}^2 \tan^2\beta,
\end{equation}
\noindent where $\beta = \frac{\pi}{2} - \alpha$ is the shifted phase offset.

Indeed, when we simulate specific parameters with $N = 32$ oscillators (Fig. \ref{fig:newbehavior_fig}C(right)), we find this transition from chiral breather to stable chiral behavior, as predicted from the infinite-$N$ analysis.

\color{black}


In order to demonstrate the flexibility and capacity of our curiosity search framework, we present an algorithmic extension which naturally incorporates human insight (Fig. \ref{fig:10population_figure}A).
We employ this human-aligned approach to explore a 100-dimensional model.

In particular, we define a 10-population Kuramoto model with global phase offset $\alpha$ (Fig. \ref{fig:10population_figure}B):
\begin{equation}\label{kuramoto_model_10population}
\dot{\theta_i^\sigma} = \omega  +  \sum_{\sigma'=1}^{10} \frac{K_{\sigma\sigma'}}{N_{\sigma'}} \sum_{j=1}^{N_{\sigma'}} \sin(\theta_j^{\sigma'} - \theta_i^\sigma - \alpha),
\end{equation}
\noindent with $\omega = 0$ and total $N = 100$ divided equally among 10 individual populations. We restrict all elements of $K_{\sigma\sigma'}$ to be positive, and further require them to sum to 1. This model is therefore 100-dimensional.
Initial exploration of the 10-population model through our non-human-aligned procedure identified several interesting behaviors, but many of the non-trivial dynamics were confined to a single population.
Having previously discovered such behaviors in the 3-population Kuramoto model, we no longer considered these behaviors to be novel, and decided to prioritize the discovery of behaviors involving multiple populations.

In order to focus sampling on multi-population behaviors, we introduce the concept of human evaluation of latent spaces, inspired by the HOLMES algorithm\cite{etcheverry2020hierarchically}.
Our human-aligned curiosity search relies on the construction of an initial latent space following the procedure outlined in Fig. \ref{fig:algorithm_schematic}.
We subsequently freeze the initial latent space and assign acceptance probabilities to each cluster, based on a human evaluation of the cluster's interest.
We now perform curiosity-driven search in a new latent space, but with sampled behaviors filtered through the initial latent space; samples are rejected or accepted based on the accepted probability of the cluster they are best associated with in the initial latent space (Fig. \ref{fig:10population_figure}C, right).

In this particular case, we scored clusters based on the presence of behaviors involving the simultaneous presence of non-trivial dynamics in multiple oscillator populations.
Therefore, the new latent space is constructed solely from sampled behaviors which have been screened through the human-evaluated initial latent space.
See Appendix \ref{human_alignment_details} for further detail.

Following human-aligned curiosity search, we construct a latent space of behaviors and cluster in that space (Fig. \ref{fig:10population_figure}C, left).
We choose representatives of each cluster and analyze the behaviors of those representatives by integrating the phase coherence curves of the whole oscillator ensemble, as well as the phase coherence curves for each of the 10 individual populations.

We find a wide variety of behaviors with non-trivial multi-population dynamics, in accordance with the human intuition we aligned our curiosity searcher with (Fig. \ref{fig:10population_figure}D). 
We can qualitatively identify certain behaviors such as breathing chimeras (behaviors 7, 14), nearly synchronized (behaviors 13, 15), and chiral phases (behavior 11).
Many of the behaviors sampled involve multiple populations overlaying in regular (behavior 1) or chaotic patterns (e.g. behaviors 3, 4, 5, 8, 9), aligning with our human-informed scoring criterion.
Even behaviors involving a single unsynchronized population can display subtle combinations of chiral and chimeric behavior; note for example the way in which the overall coherence is never complete in behavior 10, despite the periodic recurrence of coherence in all populations. \color{black}

\section*{Limitations and extensions}
While our method is successful in identifying novel phases and order parameters with minimal human effort, there are limitations on the effectiveness of our curiosity search as currently implemented. 
Many of these limitations can be traced to the geometry of the parameter space-to-behavior space map, and can be improved upon in future work.

One set of issues comes from the strength of gradients in behavior as a function of design parameters.
If the behavior is constant in a region of parameter space, then our choice to sample from locally perturbed of previously explored parameter values can result in search dynamics that is equivalent to diffusion in that region of parameter space. This local diffusion can result in a heavy dependence upon the behaviors initially sampled to seed the curiosity search. Furthermore, the problem becomes more acute as the dimension of parameter space increases.


This limitation suggests that ``messier'' physical systems, away from thermodynamic limits with sharp transitions in behaviors, may be more amenable to methods of curiosity search that operate in the space of behaviors.
There may be hints of one type of behavior hidden in examples of another behavior, and hence the curiosity search can follow a gradient, rather than relying solely on diffusion to randomly find a phase boundary.
The tradeoff of being away from a thermodynamic limits is that behaviors might not be as clearly apparent.
However, in both the case of diffusive and gradient-following dynamics, we expect that whenever a new behavior is discovered, the curiosity search algorithm will sample it with elevated frequency.

A key part of the curiosity search framework is the backmapping from behavior space to parameter space. 
Our nearest-known-neighbor choice was particularly simple, and as discussed, potentially introduces a decrease in exploration efficiency and an increased dependence on initial conditions when sampling in higher dimensions.
One possibility for decreasing the reliance on previously sampled parameters is to translate geometrical information in behavior space back into parameter space.
For example, if a target behavior sampled in behavior space lies between two points, we might sample between the two corresponding points in parameter space.
However, approaches in this vein assume that the geometry of parameter space and the geometry of behavior space are at the very least diffeomorphic.
Another possibility would be to treat the backmapping as a supervised deep learning problem, which would require iterative updates as latent space changed.

Another component of the curiosity search framework for which we made a simple choice was in latent space sampling.
While our current methodology samples latent space uniformly, it might be more efficient to explicitly sample in regions of latent space which have lower sample densities.
Another possibility is to forgo explicit sampling in latent space altogether, and instead select candidate behaviors of interest preferentially based on high reconstruction error following dimensionality reduction.
This approach is akin to novelty detection\cite{kerner2019novelty}.
A final possibility is to construct latent space in a way that more fully takes advantage of the temporal nature of the behavior, for example by using a recurrent autoencoder architecture to predict system evolution, as opposed to the vanilla convolutional VAE used here.

Finally, we made choices in the analysis of our latent space post-data collection, in particular performing agglomerative clustering on these data.
We emphasize that the clustering is a computational device to render the latent space more human-interpretable, but is not crucial for the success of the algorithm; we could have equally well have simply binned latent space.
However, to check to make sure that the choice of clustering algorithm does not significantly change the interpretation of the latent space and associated behaviors identified, we performed clustering with HDBSCAN\cite{mcinnes2017hdbscan} on all data sets generated, and were able to \textit{de novo} discover the same interesting behaviors as identified in post-processing with agglomerative clustering (Fig. \ref{fig:hdbscan_fig}) Additionally, while agglomerative clustering requires us to specify a number of expected phases, we found that HDBSCAN could automatically chose similar numbers of phases when the minimum cluster size hyperparameter was set to reasonable values (Fig. \ref{fig:hdbscan_fig}).




\section*{Discussion}

We have demonstrated that it is feasible to perform exploration of dynamical systems despite not knowing how to characterize the salient features of their behaviors (e.g., in terms of order parameters).
\color{black} This \textit{curious} exploration learns the metrics which characterize a novel system without having a pre-defined target or goal\cite{oudeyer2007intrinsic,moulin2012curiosity}. \color{black}
We achieved this curious exploration by combining the complementary strengths of active learning and dimensionality reduction; dimensionality reduction enables the iterative construction of a low-dimensional latent space of behaviors, while searching in latent space improves the efficiency of data collection.
\color{black}While active learning and dimensionality reduction have individually been applied in the context of physical systems, this approach allows us to solve a qualitatively new challenge that has yet to be confronted in a physical domain.\color{black}

We applied our method to the well-studied Kuramoto model, reproducing known behaviors in some cases and revealing novel behaviors and related novel order parameters in others. Further, the known behaviors of the canonical Kuramoto model are not thought to transfer immediately to related models such as those with excitable oscillators or for different functional forms of oscillator coupling\cite{acebron2005kuramoto,dorfler2014synchronization,kuramoto1991collective}; repeating the years of human effort that went into the canonical equations for these other models would be impractical. Our framework can be used to reveal behaviors for related models that might be of interest as accurate models of natural systems.

\color{black}Crucially, curiosity-driven search allows for the principled exploration of systems which would be intractable \textit{via} brute-force grid search.
A grid search requires $r^d$ samples, where $r$ is the resolution of each grid axis and $d$ is the dimensionality of the parameter space.
For the 3-population Kuramoto model with $d=10$ parameters we considered, implementing a coarse grid search with a resolution of $r=3$ values per dimension would require the computation of 6e4 timeseries.
For the 10-population Kuramoto model with 100 parameters, a similar resolution would require 5e47 timeseries.
The prohibitive cost of grid search underscores the contrast between the computational requirements of optimization versus exploration; while optimization requires only a fraction of the exhaustive number of evaluations required for exploration, optimization is only possible if an objective function exists to guide the search.
\color{black}



While we applied curiosity search to a canonical but \emph{in silico} model of a complex system, our algorithm can instead directly interface with a physical system by taking control of experimental knobs. This direction will allow for discovering functional behaviors that exploit unmodeled or unexpected effects in experimental systems such as non-linearities\cite{gatti2019some} or feedbacks. Much like reservoir computing\cite{tanaka2019recent} or model-free control\cite{fliess2013model}, our work here gives a systematic way of revealing behaviors that exploit complex unmodellable effects, rather than discovering them through serendipity. However, questions of time and resource cost of experimental iterations and the effectiveness of our method with only partial observations remain to be explored.

Natural applications along these lines include active matter systems with spatial structure. Recent experimental advances increasingly allow for the control of activity\cite{zhang2021spatiotemporal,ross2019controlling} and particle interactions \cite{wang2021emergent,bauerle2018self} in a space-time dependent manner, allowing for detailed density and orientation dependent motility. 
These experimental methods have opened up complex high-dimensional spatiotemporal design spaces; since order parameters are typically not available \emph{a priori} for these systems, the methods in this work might provide exciting opportunities for revealing novel behaviors.

\color{black}
An appealing feature of curiosity search is that it admits a natural way for aligning search with human intuition, a key concern in many domains of machine learning\cite{ouyang2022training}.
Crucially, alignment is accomplished without requiring explicit instructions on which areas of parameter space should be explored.
We anticipate that alignment methods will be particularly important in exploration of experimental systems with potentially wide arrays of behavior classes with features at multiple scales.
\color{black}

\textbf{Acknowledgments.}
The authors thank Michel Fruchart, Maciej Koch-Janusz, Ming Han, William Irvine, Sidney Nagel, Tom Witten, and the Flowers Lab at INRIA, for discussions. This work was supported by the Chicago Materials Research Center / MRSEC through NSF-DMR 1420709, and was in part completed with resources provided by the University of Chicago’s Research Computing Center. AM acknowledges support from the Simons Foundation.

\bibliographystyle{aiaa}
\bibliography{curiosity.bib,paperpile.bib}

\begin{thebibliography}{10}
\newcommand{\enquote}[1]{``#1''}

\bibitem{ross2019controlling}
Ross, T.~D., Lee, H.~J., Qu, Z., Banks, R.~A., Phillips, R., and Thomson, M.,
  \enquote{Controlling organization and forces in active matter through
  optically defined boundaries,} {\em Nature\/}, Vol.~572, No. 7768, 2019,
  pp.~224--229.

\bibitem{volpe2011microswimmers}
Volpe, G., Buttinoni, I., Vogt, D., K{\"u}mmerer, H.-J., and Bechinger, C.,
  \enquote{Microswimmers in patterned environments,} {\em Soft Matter\/},
  Vol.~7, No.~19, 2011, pp.~8810--8815.

\bibitem{buttinoni2012active}
Buttinoni, I., Volpe, G., K{\"u}mmel, F., Volpe, G., and Bechinger, C.,
  \enquote{Active Brownian motion tunable by light,} {\em Journal of Physics:
  Condensed Matter\/}, Vol.~24, No.~28, 2012, pp.~284129.

\bibitem{zhang2021spatiotemporal}
Zhang, R., Redford, S.~A., Ruijgrok, P.~V., Kumar, N., Mozaffari, A., Zemsky,
  S., Dinner, A.~R., Vitelli, V., Bryant, Z., Gardel, M.~L., et~al.,
  \enquote{Spatiotemporal control of liquid crystal structure and dynamics
  through activity patterning,} {\em Nature materials\/}, Vol.~20, No.~6, 2021,
  pp.~875--882.

\bibitem{bauerle2018self}
B{\"a}uerle, T., Fischer, A., Speck, T., and Bechinger, C.,
  \enquote{Self-organization of active particles by quorum sensing rules,} {\em
  Nature communications\/}, Vol.~9, No.~1, 2018, pp.~1--8.

\bibitem{wang2021emergent}
Wang, G., Phan, T.~V., Li, S., Wombacher, M., Qu, J., Peng, Y., Chen, G.,
  Goldman, D.~I., Levin, S.~A., Austin, R.~H., et~al., \enquote{Emergent
  field-driven robot swarm states,} {\em Physical review letters\/}, Vol.~126,
  No.~10, 2021, pp.~108002.

\bibitem{dai2020efficient}
Dai, C. and Glotzer, S.~C., \enquote{Efficient phase diagram sampling by active
  learning,} {\em The Journal of Physical Chemistry B\/}, Vol.~124, No.~7,
  2020, pp.~1275--1284.

\bibitem{whitelam2021neuroevolutionary}
Whitelam, S. and Tamblyn, I., \enquote{Neuroevolutionary learning of particles
  and protocols for self-assembly,} {\em Physical review letters\/}, Vol.~127,
  No.~1, 2021, pp.~018003.

\bibitem{ferguson2022data}
Ferguson, A.~L. and Brown, K.~A., \enquote{Data-driven design and autonomous
  experimentation in soft and biological materials engineering,} {\em Annual
  Review of Chemical and Biomolecular Engineering\/}, Vol.~13, 2022.

\bibitem{shmilovich2020discovery}
Shmilovich, K., Mansbach, R.~A., Sidky, H., Dunne, O.~E., Panda, S.~S., Tovar,
  J.~D., and Ferguson, A.~L., \enquote{Discovery of self-assembling
  $\pi$-conjugated peptides by active learning-directed coarse-grained
  molecular simulation,} {\em The Journal of Physical Chemistry B\/}, Vol.~124,
  No.~19, 2020, pp.~3873--3891.

\bibitem{mohr2022data}
Mohr, B., Shmilovich, K., Kleinw{\"a}chter, I.~S., Schneider, D., Ferguson,
  A.~L., and Bereau, T., \enquote{Data-driven discovery of
  cardiolipin-selective small molecules by computational active learning,} {\em
  Chemical Science\/}, Vol.~13, No.~16, 2022, pp.~4498--4511.

\bibitem{grizou2020curious}
Grizou, J., Points, L.~J., Sharma, A., and Cronin, L., \enquote{A curious
  formulation robot enables the discovery of a novel protocell behavior,} {\em
  Science advances\/}, Vol.~6, No.~5, 2020, pp.~eaay4237.

\bibitem{oudeyer2007intrinsic}
Oudeyer, P.-Y., Kaplan, F., and Hafner, V.~V., \enquote{Intrinsic motivation
  systems for autonomous mental development,} {\em IEEE transactions on
  evolutionary computation\/}, Vol.~11, No.~2, 2007, pp.~265--286.

\bibitem{reinke2019intrinsically}
Reinke, C., Etcheverry, M., and Oudeyer, P.-Y., \enquote{Intrinsically
  Motivated Discovery of Diverse Patterns in Self-Organizing Systems,} {\em
  International Conference on Learning Representations\/}, 2019.

\bibitem{coli2022inverse}
Coli, G.~M., Boattini, E., Filion, L., and Dijkstra, M., \enquote{Inverse
  design of soft materials via a deep learning--based evolutionary strategy,}
  {\em Science advances\/}, Vol.~8, No.~3, 2022, pp.~eabj6731.

\bibitem{thiem2020emergent}
Thiem, T.~N., Kooshkbaghi, M., Bertalan, T., Laing, C.~R., and Kevrekidis,
  I.~G., \enquote{Emergent spaces for coupled oscillators,} {\em Frontiers in
  computational neuroscience\/}, Vol.~14, 2020, pp.~36.

\bibitem{carrasquilla2017machine}
Carrasquilla, J. and Melko, R.~G., \enquote{Machine learning phases of matter,}
  {\em Nature Physics\/}, Vol.~13, No.~5, 2017, pp.~431--434.

\bibitem{mcgibbon2017identification}
McGibbon, R.~T., Husic, B.~E., and Pande, V.~S., \enquote{Identification of
  simple reaction coordinates from complex dynamics,} {\em The Journal of
  Chemical Physics\/}, Vol.~146, No.~4, 2017, pp.~044109.

\bibitem{van2020classifying}
Van~Damme, R., Coli, G.~M., Van~Roij, R., and Dijkstra, M.,
  \enquote{Classifying crystals of rounded tetrahedra and determining their
  order parameters using dimensionality reduction,} {\em ACS nano\/}, Vol.~14,
  No.~11, 2020, pp.~15144--15153.

\bibitem{vaddi2022autonomous}
Vaddi, K., Chiang, H.~T., and Pozzo, L.~D., \enquote{Autonomous retrosynthesis
  of gold nanoparticles via spectral shape matching,} {\em Digital
  Discovery\/}, Vol.~1, No.~4, 2022, pp.~502--510.

\bibitem{gilpin2021chaos}
Gilpin, W., \enquote{Chaos as an interpretable benchmark for forecasting and
  data-driven modelling,} {\em arXiv preprint arXiv:2110.05266\/}, 2021.

\bibitem{ricci2022phase2vec}
Ricci, M., Moriel, N., Piran, Z., and Nitzan, M., \enquote{Phase2vec: Dynamical
  systems embedding with a physics-informed convolutional network,} {\em arXiv
  preprint arXiv:2212.03857\/}, 2022.

\bibitem{baranes2013active}
Baranes, A. and Oudeyer, P.-Y., \enquote{Active learning of inverse models with
  intrinsically motivated goal exploration in robots,} {\em Robotics and
  Autonomous Systems\/}, Vol.~61, No.~1, 2013, pp.~49--73.

\bibitem{kuramoto1975self}
Kuramoto, Y., \enquote{Self-entrainment of a population of coupled non-linear
  oscillators,} {\em International Symposium on Mathematical Problems in
  Theoretical Physics: January 23--29, 1975, Kyoto University, Kyoto/Japan\/},
  Springer, 1975, pp. 420--422.

\bibitem{acebron2005kuramoto}
Acebr{\'o}n, J.~A., Bonilla, L.~L., Vicente, C. J.~P., Ritort, F., and Spigler,
  R., \enquote{The Kuramoto model: A simple paradigm for synchronization
  phenomena,} {\em Reviews of modern physics\/}, Vol.~77, No.~1, 2005, pp.~137.

\bibitem{kingma2013auto}
Kingma, D.~P. and Welling, M., \enquote{Auto-encoding variational bayes,} {\em
  arXiv preprint arXiv:1312.6114\/}, 2013.

\bibitem{abrams2008solvable}
Abrams, D.~M., Mirollo, R., Strogatz, S.~H., and Wiley, D.~A.,
  \enquote{Solvable model for chimera states of coupled oscillators,} {\em
  Physical review letters\/}, Vol.~101, No.~8, 2008, pp.~084103.

\bibitem{dempster2020rocket}
Dempster, A., Petitjean, F., and Webb, G.~I., \enquote{ROCKET: exceptionally
  fast and accurate time series classification using random convolutional
  kernels,} {\em Data Mining and Knowledge Discovery\/}, Vol.~34, No.~5, 2020,
  pp.~1454--1495.

\bibitem{cho2017stable}
Cho, Y.~S., Nishikawa, T., and Motter, A.~E., \enquote{Stable chimeras and
  independently synchronizable clusters,} {\em Physical review letters\/},
  Vol.~119, No.~8, 2017, pp.~084101.

\bibitem{nicolaou2019multifaceted}
Nicolaou, Z.~G., Eroglu, D., and Motter, A.~E., \enquote{Multifaceted dynamics
  of Janus oscillator networks,} {\em Physical Review X\/}, Vol.~9, No.~1,
  2019, pp.~011017.

\bibitem{zhang2020critical}
Zhang, Y., Nicolaou, Z.~G., Hart, J.~D., Roy, R., and Motter, A.~E.,
  \enquote{Critical switching in globally attractive chimeras,} {\em Physical
  Review X\/}, Vol.~10, No.~1, 2020, pp.~011044.

\bibitem{abrams2004chimera}
Abrams, D.~M. and Strogatz, S.~H., \enquote{Chimera states for coupled
  oscillators,} {\em Physical review letters\/}, Vol.~93, No.~17, 2004,
  pp.~174102.

\bibitem{fruchart2021non}
Fruchart, M., Hanai, R., Littlewood, P.~B., and Vitelli, V.,
  \enquote{Non-reciprocal phase transitions,} {\em Nature\/}, Vol.~592, No.
  7854, 2021, pp.~363--369.

\bibitem{ott2008low}
Ott, E. and Antonsen, T.~M., \enquote{Low dimensional behavior of large systems
  of globally coupled oscillators,} {\em Chaos: An Interdisciplinary Journal of
  Nonlinear Science\/}, Vol.~18, No.~3, 2008, pp.~037113.

\bibitem{etcheverry2020hierarchically}
Etcheverry, M., Moulin-Frier, C., and Oudeyer, P.-Y., \enquote{Hierarchically
  organized latent modules for exploratory search in morphogenetic systems,}
  {\em Advances in Neural Information Processing Systems\/}, Vol.~33, 2020,
  pp.~4846--4859.

\bibitem{kerner2019novelty}
Kerner, H.~R., Wellington, D.~F., Wagstaff, K.~L., Bell, J.~F., Kwan, C., and
  Amor, H.~B., \enquote{Novelty detection for multispectral images with
  application to planetary exploration,} {\em Proceedings of the aaai
  conference on artificial intelligence\/}, Vol.~33, 2019, pp. 9484--9491.

\bibitem{mcinnes2017hdbscan}
McInnes, L., Healy, J., and Astels, S., \enquote{hdbscan: Hierarchical density
  based clustering.} {\em J. Open Source Softw.\/}, Vol.~2, No.~11, 2017,
  pp.~205.

\bibitem{moulin2012curiosity}
Moulin-Frier, C. and Oudeyer, P.-Y., \enquote{Curiosity-driven phonetic
  learning,} {\em 2012 IEEE international conference on development and
  learning and epigenetic robotics (ICDL)\/}, IEEE, 2012, pp. 1--8.

\bibitem{dorfler2014synchronization}
D{\"o}rfler, F. and Bullo, F., \enquote{Synchronization in complex networks of
  phase oscillators: A survey,} {\em Automatica\/}, Vol.~50, No.~6, 2014,
  pp.~1539--1564.

\bibitem{kuramoto1991collective}
Kuramoto, Y., \enquote{Collective synchronization of pulse-coupled oscillators
  and excitable units,} {\em Physica D: Nonlinear Phenomena\/}, Vol.~50, No.~1,
  1991, pp.~15--30.

\bibitem{gatti2019some}
Gatti, G., Brennan, M., and Tang, B., \enquote{Some diverse examples of
  exploiting the beneficial effects of geometric stiffness nonlinearity,} {\em
  Mechanical Systems and Signal Processing\/}, Vol.~125, 2019, pp.~4--20.

\bibitem{tanaka2019recent}
Tanaka, G., Yamane, T., H{\'e}roux, J.~B., Nakane, R., Kanazawa, N., Takeda,
  S., Numata, H., Nakano, D., and Hirose, A., \enquote{Recent advances in
  physical reservoir computing: A review,} {\em Neural Networks\/}, Vol.~115,
  2019, pp.~100--123.

\bibitem{fliess2013model}
Fliess, M. and Join, C., \enquote{Model-free control,} {\em International
  Journal of Control\/}, Vol.~86, No.~12, 2013, pp.~2228--2252.

\bibitem{ouyang2022training}
Ouyang, L., Wu, J., Jiang, X., Almeida, D., Wainwright, C., Mishkin, P., Zhang,
  C., Agarwal, S., Slama, K., Ray, A., et~al., \enquote{Training language
  models to follow instructions with human feedback,} {\em Advances in Neural
  Information Processing Systems\/}, Vol.~35, 2022, pp.~27730--27744.

\end{thebibliography}

\appendix

\section{simulation details}
\label{simulation_details}

Our primary goal is to investigate and discover novel dynamical behaviors in variants of the Kuramoto model, all of which can be compactly written in the following form:
\begin{equation}\label{general_kuramoto}
\dot{\theta_i} = \omega_i  +  \sum_{j=1}^N K_{ij}  \sin(\theta_j - \theta_i - \alpha),
\end{equation}
\noindent where $K_{ij}$ is the matrix of couplings between oscillators, $\alpha$ is a global phase offset, the intrinsic frequencies $\omega_i$ are (potentially) drawn from a distribution, and $N$ is the number of oscillators.
In all figures, $N$ is generally on the order of 30, and $\omega_i = 0$ except for in the uniformly-connected model considered in Figure 2. 
In the uniformly-connected model, $\omega_i$ is drawn from the distribution $\mathcal{N}(0,.1)$.
Additionally in the uniformly-connected model, $\alpha = 0$.

In order to investigate these dynamical systems, we need to integrate Eq. \ref{general_kuramoto} for specific $K_{ij}$ and $\alpha$, for which we use the SciPy odeint function.
We supply a regular time grid of domain $[0,750]$ with step size $dt = .05$.
Initial oscillator phases are drawn from a uniform distribution from $[0, 2\pi]$.

Traditionally, the output of these models has been investigated using a ``phase coherence'' order parameter, which can be defined as $|\frac{1}{N}\sum_{j=1}^N e^{i\theta_j}|$.
There is also the associated complex phase $\arg\left[\frac{1}{N}\sum_{j=1}^N e^{i\theta_j}\right]$, but we focus on this less in our current work.

Instead of interpreting and processing the output of our dynamical system with these traditional metrics, we allow unsupervised dimensionality reduction techniques (see Appendix \ref{dimreduction_details}) to extract the relevant order parameters.
In order to pass the raw output of the dynamical systems integration to the dimensionality reduction technique, we sample the last quarter of the raw output timeseries in $7$ evenly-spaced intervals.
At each sample, we use the NumPy arctan2 function to compute the mean oscillator phase, and then compute the distribution of sines of oscillator angles relative to the mean.
The sine values are subsequently binned in the range $[-1,1]$ with 7 bins, and the histogram is normalized by the number of oscillators.
Having done this for 7 timepoints, we have transformed our raw dynamical output into a $7 \times 7$ greyscale image, which is the input to a dimensionality reduction technique.
Note that by mean-centering at each time point and binning, we imposed invariance to oscillation index, as well as invariance to global rotations.
In the context of coupled oscillator models, these assumptions seem relatively benign, but may not be appropriate for other systems.

\section{active learning details}\label{active_learning_details}

Having described how we compute the behavior for a given set of Kuramoto model parameters, we can now turn to the active learning procedure by which we sample model parameters.
All our explorations are seeded by collecting 200 (uniformly-connected and chimera models) or 800 (3-population model) samples randomly throughout parameter space.
Each parameter space axis has an upper and lower bound: for phase offsets, this is $[0,\frac{\pi}{2}]$; for oscillator couplings, this is $[0,1]$ in the chimera and 3-population models, and $[0,2]$ in the uniformly-connected model.

The parameters are then used to integrate the dynamical equation described in Appendix \ref{simulation_details}.
We note that, for the chimera and 3-population models, initial oscillator phases are sampled once at the beginning of the active learning procedure, and subsequently fixed for the duration of the exploration. 
In contrast, for the fully-connected model, the initial oscillator phases and the intrinsic frequencies are resampled with each new parameter selected.
The output of these initial simulations are passed through a dimensionality reduction technique (Appendix \ref{dimreduction_details}). If the employed technique requires training, training is also performed before the dynamical behaviors are converted to their latent space representations.

We have now initialized our active learning exploration, by creating a collection of tuples containing all our relevant information: (parameters, dynamic behaviors, latent space representation).
Following initialization, we now select a ``target'' behavior in latent space that we wish to explore.
While there are many possible options for performing this latent space sampling, we pick a particularly simple one; we construct the hyperrectangle that contains all the currently sampled latent space representations, and then uniformly sample within that hyperrectangle.

With this target behavior in hand, we now seek a point in parameter space that will ideally lead us to this target point in latent space.
Again, there are many possible options for implementing this parameter point selection, and we choose a simple one.
In this case, we return to our dictionary of all previously sampled parameters, and select the parameter whose latent space representation is closest to our target.
We then ``mutate'' this selected parameter by adding a random amount along each parameter space axis. 
The magnitude of this random step along each parameter space axis is bounded by 10\% (uniformly-connected and chimera models) or 20\% (3-population) of the allowed domain length of that axis. Within these bounds, the step length is sampled uniformly.
If the mutated parameter falls outside the lower or upper bound of any of the parameter axes, we resample the mutation until the mutated parameter falls within the allowed ranges.

We can continue this process, collecting more (parameter, dynamic behavior, latent space representation)-tuples.
We can also iteratively train the associated dimensionality reduction techniques with this newly collected data.
For the uniformly-connected and chimera model explorations, we updated the dimensionality reduction technique once every 100 samples collected, until we had a total of 1600 collected samples, inclusive of the initial samples.
For the 3-population model exploration, we updated the dimensionality reduction technique once every 400 samples collected, until we collected a total of 4000 samples. 
When we select samples for training, we utilize 50\% from the most recent samples, and 50\% randomly chosen from the previous samples.

\section{dimensionality reduction details}
\label{dimreduction_details}

\begin{figure*}
\begin{centering}
\includegraphics[width=\linewidth]{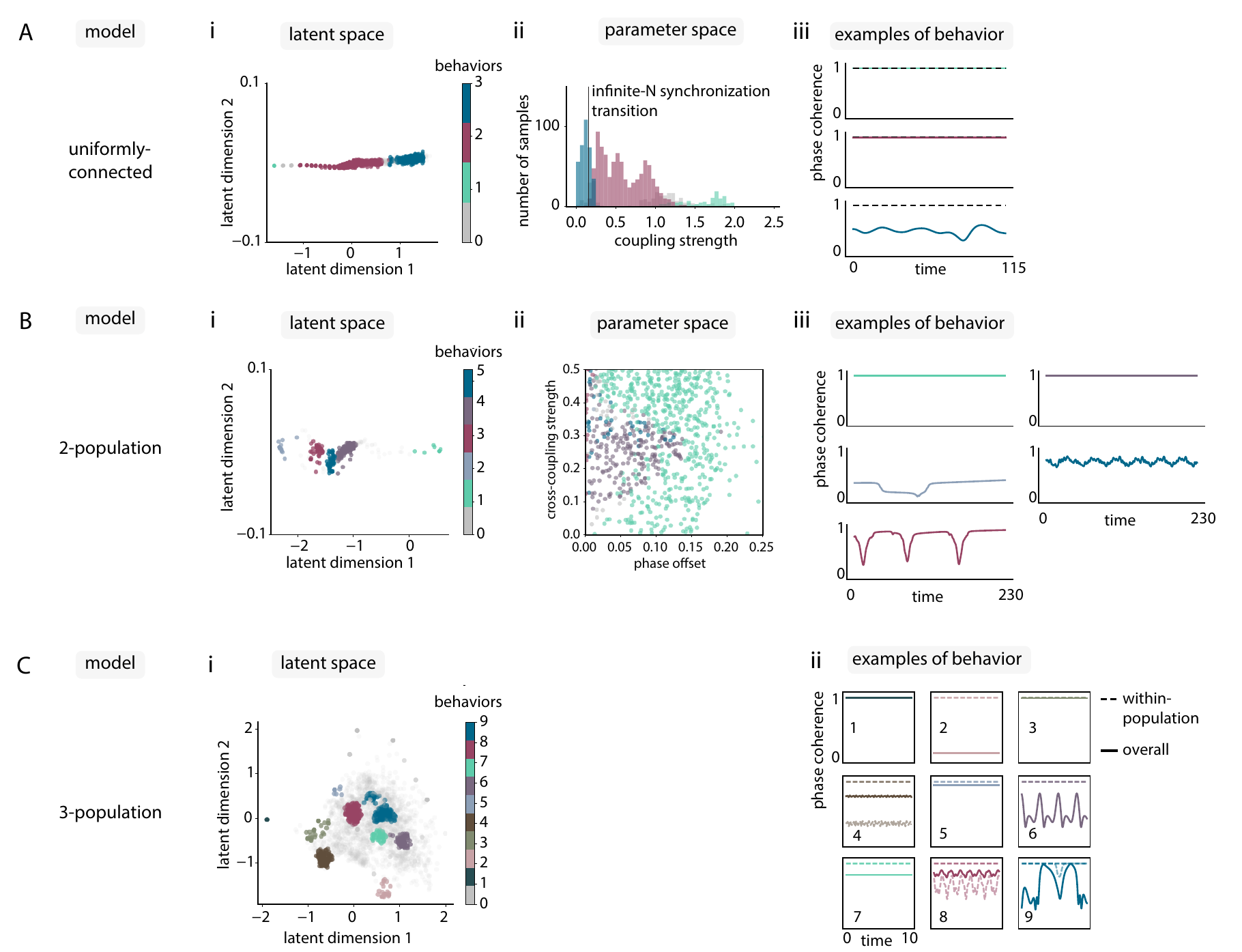}
\par\end{centering}
\centering{}\caption{Clustering with HDBSCAN on curiosity search data reveals novels dynamical behaviors in Kuramoto model variants. (A) Clustering with HDBSCAN in the latent space of the uniformly-connected model identifies 3 clusters (i), which map to regimes of low, intermediate, and high synchronization in parameter space (ii, iii). (B) Clustering with HDBSCAN in the latent space of the chimera model identifies 5 clusters (i) which can roughly distinguish between chimeric and fully-synchronized phases (ii, iii). (C) Clustering with HDBSCAN in the latent space of the 3-population model identifies 9 clusters (i), which reveal similar behaviors to those identified in the main text (ii), including the chiral breather behavior (behavior 6).}
\label{fig:hdbscan_fig}
\end{figure*}

For each coupled oscillator model exploration, we consider four dimensionality reduction techniques in the paper: a convolutional variational autoencoder 
(VAE), a random VAE, PCA, and random projection.
All neural network code was run using PyTorch, and the linear models were implemented with Scikit-learn.

The convolutional VAE has a relatively simple encoder 
architecture: 1. a 2D convolutional layer with 2-4 filters, followed by ReLU activation and flattening; and then 2. a fully-connected layer into a latent space of dimension 2-4.
The decoder follows analogously: 1. a fully-connected layer which expands from the latent space, followed by unflattening and ReLU activation; and then 2. a transposed 2D convolution followed by sigmoid activation.
For the chimera and uniformly-connected models, we use 2 filters and 2 latent dimensions, whereas for the 3-population model we use 4 filters and 4 latent dimensions.
Every time the VAE is trained, it is trained on a batch size of 200 or 800 (chosen as described in Appendix \ref{active_learning_details}) for 2000 epochs. We train with ADAM, using a learning rate of $1\mathrm{e}-3$ and weight decay $1\mathrm{e}-5$.
Weights are initialized with PyTorch standard initialization, which in linear and convolutional layers with ReLU activation is he normalization.
The random VAE is constructed in exactly the same architecture and initialization as the corresponding trained VAE, but is never trained over the course of the active learning.

PCA is performed using the Scikit-learn PCA method, and we utilize the same number of dimensions as used for the VAEs in order to construct the PCA latent space. Random projection is performed using the Scikit-learn GaussianRandomProjection function, projected onto the same number of dimensions as used for the VAE latent spaces.

\section{clustering and behavior example selection}
\label{sec:clustering_methods}

In order to interpret latent space, we employ agglomerative clustering as implemented in the 
scikit-learn AgglomerativeClustering function with Ward linkage. We choose 3, 6, and 10 cluster for Figures 2, 3, and 4 respectively. In order to gain a qualitative understanding of these clusters, we select the sample closest to the cluster median (in latent space) and then assess the resulting dynamics for the corresponding point in parameter space. The precise values of the selected parameters are presented in Supplemental Table 1. Note that in the 3-population model, the listed couplings are normalized to 1 before being used as model input.

To evaluate the robustness of our clustering-related conclusions, we also perform clustering using HDBSCAN\cite{mcinnes2017hdbscan} on the same dataset analyzed in the main figures. 
In contrast to agglomerative clustering, HDBSCAN does not require the number of desired clusters as a hyperparameter. Hence as a first check on the consistency of our results, we checked whether our chosen agglomerative cluster numbers could be reproduced with reasonable values of the HDBSCAN hyperparameter min\_cluster\_size, which sets the minimum allowable cluster size. 
For the uniformly-connected model, we were most interested in coarse features, so we set a minimum cluster size of 100 (out of 1600 samples) to select 3 clusters (Fig. \ref{fig:hdbscan_fig}Ai). 
For the chimera model, we were more interested in fine-grained distinctions, so we set a minimum cluster size of 20 (out of 1600 samples) to select 5 clusters (Fig. \ref{fig:hdbscan_fig}Bi). 
For the 3-population model, we were again interested in fine-grained distinctions, so we set a minimum cluster size of 80 (out of 4000 samples) to select 9 clusters (Fig. \ref{fig:hdbscan_fig}Ci).

As a second check on the robustness of our results, we asked whether we could identify the same interesting phases we found using agglomerative clustering in an HDBSCAN-derived clustering.
Note that HDBSCAN identifies a category of points as noise, which in all panels we color as grey and label as behavior 0.
For the uniformly-connected model, we again recovered the low (behavior 3), intermediate (behavior 2), and fully-synchronized (behavior 1) regimes (Fig. \ref{fig:hdbscan_fig}Aiii).
For the chimera model, we were able to distinguish chimeric (behaviors 4 and 5) from fully-synchronized regimes (behavior 1) (Fig. \ref{fig:hdbscan_fig}Bii).
There is also some distinction between breathing and stable chimeras, though the splitting is less clean than in the agglomerative clustering case.
This suggests that the latent space is capable of distinguishing between the two chimera variants, but that this particular clustering is slightly too coarse to cleanly find the dividing line.
Finally, we again discover a similar range of behaviors in the 3-population model as under the agglomerative clustering analysis: fully-synchronized (behaviors 1, 3), chimera (behaviors 4, 8), chiral (behaviors 7, 5), anti-aligned (behavior 2), chiral breathers (behavior 6), as well as behaviors with some combination of chimeric and chiral characteristics (behavior 9) (Fig. \ref{fig:hdbscan_fig}Ciii).

\section{algorithm performance metrics}

\subsection{ideal sampling comparison incorporating prior model knowledge}
\label{sec:ideal_measure}

To assess the performance of the various parameter exploration schemes outlined in the main text, we want to quantify the quality of the sampling distributions they generate in parameter space. In particular, an ideal benchmarking measure would compare curiosity searches against a known, desired, sampling distribution.

In the uniformly-connected Kuramoto model, we have prior knowledge about the various phases we expect to see.
In the infinite-N limit, we know that there are two well-defined phases; a fully incoherent phase in which the Kuramoto order parameter $r=0$, and above a critical coupling $K_c$ a synchronized parameter regime in which $r > 0$.
With this prior, we would ideally like our sampling to be evenly distributed, with half the samples coming from above $K_c$ and half below.
However, our simulations are performed with finite N, and hence we should not expect such cleanly delineated phases.

Instead, we define an ideal sampling by the computation of $r$ in our simulations as a function of $K$.
We select three regimes: one phase for which $r < .3$, with the corresponding parameter values $K < .13$, one for which $r > .95$ and hence $K > .34$, and the phase intermediate to those two.
We posit that the ideal sampling distribution should be evenly distributed between these three regimes in parameter space.

To quantitatively make the comparison between our sampled distributions and the ideal distribution, we use the scipy.special.kl\_div function to compute the KL divergence $D_{KL}(\text{sampled}|\text{ideal})$ between the two distributions.
This is the number we use as our performance metric for a particular sampling distribution.

We also have ground truth knowledge in the case of the chimera model, allowing us to compare the quality of sampling distributions analogously.
We identify three phases as stable chimeras, breathing chimeras, and synchronized phases outside of the previous two regions.
These phases and their boundaries in parameter space were identified in Abrams et al.\cite{abrams2008solvable};
while other phases might in principle exist, we do not incorporate this possibility into the analysis.

Based on the work of Abrams et al., we can estimate these boundaries using the shapely python package to define points that lie within the triangle $[(0,0), (0,0.2679), (0.2239,0.3372)]$ to be stable chimeras; points that lie within the triangle $[(0,0.2679), (0.2239,0.3372), (0,0.5)]$ to be breathing chimeras; and points outside these triangles to be synchronized phases.
Given this procedure of computing phases in parameter space, we follow the same procedure as we did for the uniformly-connected Kuramoto model; we assume ideal sampling is even across the three phases, and then compute the KL divergence between each sampling distribution and ideal sampling.

\begin{figure*}
\begin{centering}
\includegraphics[width=\linewidth]{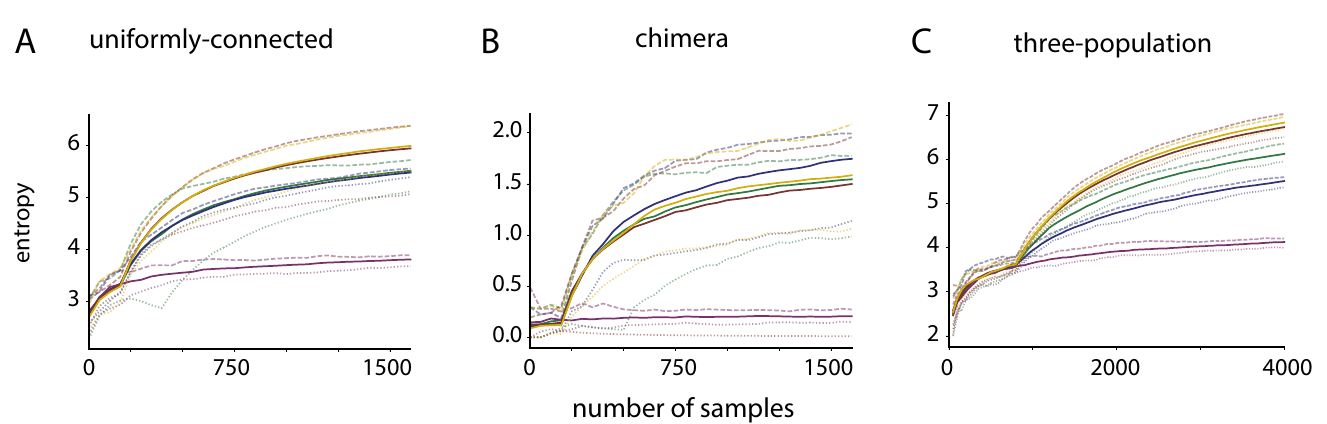}
\par\end{centering}
\centering{}\caption{Curiosity search generates higher entropy sampling distributions compared to random parameter space sampling. Higher entropy indicates more uniform sampling in the autoencoder latent space of the 
(A) uniformly-connected, (B) chimera, and (C) three-population models.}
\label{fig:entropy_fig}
\end{figure*}

\subsection{model-agnostic diversity and entropy measures}
\label{sec:diversity_measure}


Running the curiosity algorithm with different dimensionality reduction techniques (see Appendix \ref{dimreduction_details}) generates different distributions of samples.
We would like to compare the performance of each technique in terms of generating a more diverse collection of samples, relative to a random parameter sampling baseline. 

In order to compare sampling distributions, we must construct a measure of diversity. For the fully-connected and chimera models, we have prior knowledge of how many phases exist, so we can simply characterize how well the different techniques sample these known phases. 
However, for the 3-family model, we do not have this prior knowledge. Therefore, we want to construct a diversity measure which does not depend on complete prior knowledge. 

One way to do this is construct a measure which captures the diversity of sampling in latent space, as latent space is a representation of the system behaviors.
However, each dimensionality reduction technique constructs a different latent space.
We make the assumption that the trained autoencoder latent space is the most tailored latent space, and so to create comparable representations, we run each distribution of collected dynamical behaviors through the same trained autoencoder.
We then normalize the autoencoder latent-space based on the full collection of latent-space representations from all distributions. 

Finally, to calculate a measure of diversity for a distribution of samples, we divide each dimension of the normalized latent space into 80 bins. 
This divides the latent space into hyper-cubes, the size and number of which is determined by the bin number. 
Each latent space value fits into one of these cubes.
We subsequently define our measure of diversity for our sample distribution as the number of unique cubes occupied by all samples. 
Note that the number of samples which lie in each cube is not considered, only the number of unique cubes filled.

From this construction, we can similarly measure the entropy of the various latent space sampling distributions. 
Higher entropy indicates more uniform coverage of latent space. 
For each model we investigated, we found that the entropy of latent space samples from each dimensionality reduction technique was greater than the entropy of the behavior distributions generated from random parameter sampling (Fig. \ref{fig:entropy_fig}).

\subsection{temporal sampling of fully-connected model}
\label{sec:temporal_sampling}

To construct a representation of how the fully-connected model sampled over time with a periodically retrained autoencoder, we divided the parameter space (coupling strength) into 8 bins, and then normalized each bin individually from 0 to 100 percent. Since the samples were saved in the order in which they were collected, dividing the array of samples into quartiles is equivalent to dividing it into four sequential temporal bins. We plotted what percent of the samples in each parameter space bin lay in each quartile. This method of visualization shows where the algorithm preferentially sampled over time.

\color{black}
\section{human-aligned curiosity search}\label{human_alignment_details}

Human intuition can naturally be incorporated into the curiosity search framework presented in Fig. \ref{fig:algorithm_schematic} with a simple, potentially iterative modification, following the spirit of Ref. \cite{etcheverry2020hierarchically}.
At arbitrary times within the curiosity search loop, latent space can be frozen, and a human observer can score the behaviors presented in the various parts of that frozen latent space for human interest.
This scoring can in turn be converted to acceptance probabilities.
Subsequent curiosity sampling can run all produced behaviors through that latent space, and the sample is rejected or accepted according to those human-derived acceptance probabilities.
In principle, this process of latent space freezing and human-alignment can be continued iteratively, with freezing occurring on human-set intervals or triggered autonomously by quantitative criteria\cite{etcheverry2020hierarchically}.

We demonstrate this algorithmic variant in the context of a Kuramoto model with 10 populations.
We first performed a naive curiosity search with an autoencoder with 8 latent dimensions and 8 filters in the initial convolutional layer. Input to the autoencoder was constructed as before, but with 13 bins for computing oscillator phase space density as opposed to the original 7. These densities were computed at 13 timepoints, as opposed to 7.
In this initial search, collected a total of 4000 samples and updated the dimensionality reduction technique once every 400 samples. 

We found that the algorithm identified several interesting behaviors, but many of the non-trivial dynamics were confined to a single population.
Having already seen such behaviors in the 3-populaton Kuramoto model, we no longer considered these behaviors to be novel, and decided to prioritize the discovery of behaviors with non-trivial dynamics occurring in multiple populations simultaneously.
Therefore, we clustered our naive latent space using the scikit-learn KMeans function with 15 clusters and default hyperparameters.
We visualized the behaviors present at the sample closest to the mean of each cluster, and assigned an acceptance probability to that cluster; if the cluster was synchronized or nearly synchronized, it received an acceptance probability of 0; for non-trival dynamics involving a single-population, we assigned a probability of .5; and for non-trivial dynamics involving multiple populations, we assigned a probability of 1.

We now began a human-aligned portion of our search.
We initialize with the samples from clusters in the initial, naive run which received an acceptance probability of 1.
Our new autoencoder is initialized with weights from the trained autoencoder saved at the end of the naive run. 
Every time we sample a new behavior, we run it through the old, naive autoencoder, and determine the cluster of the old latent space the new sample lies by calling the KMeans predict method.
Based on the acceptance probability of the predicted cluster, that sample is accepted or rejected.
If the sample is accepted, the algorithm proceeds normally. 
If the sample is rejected, a new sample is taken in parameter space. 
We collect 3200 (accepted) samples following this procedure. The new autoencoder was retrained for 2000 epochs every 400 samples.
Details concerning integration of the dynamical system are identical to those presented in Appendix \ref{simulation_details}.\color{black}

\end{document}